\documentclass[useAMS,usenatbib]{mn2e}
\usepackage{graphicx,amssym,color}
\newcommand{\Mpch}{\mbox{ $h^{-1}$ Mpc}}

\newcommand{\be}{\begin{equation}}
\newcommand{\ee}{\end{equation}}

\def\ltsima{$\; \buildrel < \over \sim \;$}
\def\simlt{\lower.5ex\hbox{\ltsima}}
\def\gtsima{$\; \buildrel > \over \sim \;$}
\def\simgt{\lower.5ex\hbox{\gtsima}}
\def\la{$\langle$}

\title{The Mass-Concentration-Redshift Relation of Cold Dark Matter Halos}

\author[Ludlow et al.] {\parbox{18cm}{
Aaron D. Ludlow$^{1,\star}$,
Julio F. Navarro$^{2}$,
Ra\'ul E. Angulo$^{3}$,
Michael Boylan-Kolchin$^{4}$,
Volker~Springel$^{5,6}$,
Carlos Frenk$^{7}$,
Simon D. M. White$^{8}$
}\vspace{0.3cm}\\
$^{1}${Argelander-Institut f\"{u}r Astronomie, Auf dem H\"{u}gel 71,
  D-53121 Bonn, Germany}\\
$^{2}${Senior CIfAR Fellow, Dept. of Physics and Astronomy, University of
  Victoria, Victoria, BC, V8P 5C2, Canada}\\
$^{3}${Centro de Estudios de F\'isica del Cosmos de Arag\'on,  
  Plaza San Juan 1, Planta-2, 44001, Teruel, Spain}\\
$^{4}${Department of Astronomy and Joint Space-Science Institute, University of Maryland,  
  College Park, MD, 20742-2421, USA}\\
$^{5}${Heidelberg Institute for Theoretical Studies,
  Schloss-Wolfsbrunnenweg 35, 69118 Heidelberg, Germany}\\
$^{6}${Zentrum f\"{u}r Astronomie der Universit\"{a}t Heidelberg, ARI,
  M\"{o}nchhofstr. 12-14, 69120 Heidelberg, Germany}\\
$^{7}${Institute for Computational Cosmology, Dept. of Physics, Univ. of
  Durham, South Road, Durham  DH1 3LE, UK}\\
$^{8}${Max-Planck-Institut f\"{u}r Astrophysik,
  Karl-Schwarzschild-Stra\ss{}e 1, 85740 Garching bei M\"{u}nchen, Germany}\\
}

\begin{document}

\maketitle 

\begin{abstract}
  We use the Millennium Simulation series to investigate the mass and
  redshift dependence of the concentration of equilibrium cold dark matter (CDM)
  halos. We extend earlier work on the relation between halo mass profiles
  and assembly histories to show how the latter may be used to predict
  concentrations for halos of all masses and at any redshift.  Our
  results clarify the link between concentration and the ``collapse
  redshift'' of a halo as well as why concentration depends on mass
  and redshift solely through the dimensionless ``peak height'' mass
  parameter, $\nu(M,z)=\delta_{\rm crit}(z)/\sigma(M,z)$. We combine
  these results with analytic mass accretion histories to extrapolate
  the $c(M,z)$ relations to mass regimes difficult to reach through
  direct simulation. Our model predicts that, at given $z$, $c(M)$
  should deviate systematically from a simple power law at high
  masses, where concentrations approach a constant value, and at low
  masses, where concentrations are substantially lower than expected
  from extrapolating published empirical fits. This correction may reduce 
  the expected self-annihilation boost factor from substructure by about 
  one order of magnitude. The model also
  reproduces the $c(M,z)$ dependence on cosmological parameters
  reported in earlier work, and thus provides a simple and robust
  account of the relation between cosmology and the
  mass-concentration-redshift relation of CDM halos.
\end{abstract}

\renewcommand{\thefootnote}{\fnsymbol{footnote}}
\footnotetext[1]{E-mail: aludlow@astro.uni-bonn.de} 

\section{Introduction}
\label{SecIntro}

It is now well established that the equilibrium density profile of cold
dark matter (CDM) halos is nearly self-similar, and may be well
approximated by scaling a simple formula \citep[][hereafter NFW]{Navarro1996,Navarro1997}:
\begin{equation}
  {\rho (r)\over \rho_{\rm crit}} = \frac{\delta_c}{(r/r_s) \, (1+r/r_s)^2}.  \label{EqNFW}
\end{equation}
This profile is referred to in the literature as the ``NFW profile''
and is characterized by a scale radius, $r_s$, and an overdensity,
$\delta_c$. These scaling parameters may also be expressed in terms of
the halo virial\footnote{We define the virial parameters of a halo as
  those measured within a sphere centered at the potential minimum
  that encloses a mean density $200\times$ the critical density for
  closure, $\rho_{\rm crit}(z)=3H(z)^2/8\pi G$, and label them with a
  ``200'' subscript. For example, $r_{200}$ and $M_{200}$ are the
  halo's virial radius and mass, respectively.} mass, $M_{200}$, and a
dimensionless ``concentration'' parameter, $c=r_{200}/r_s$, defined as
the ratio between the virial and scale radius of a halo. The scale
radius indicates where the logarithmic slope of the profile has the
isothermal value of $-2$, and therefore, we shall hereafter use indistinctly
$r_{-2}$ or $r_s$ to denote this radius. The concentration is an alternative
measure of the halo characteristic density, and at a given halos mass the 
two are related by
\begin{equation}
  \delta_c={200 \over 3} {c^3 \over [\ln(1+c)-c/(1+c)]}.
  \label{EqcDeltac}
\end{equation} 
As noted in the original NFW papers, mass and concentration are
strongly correlated, albeit with considerable scatter. 

Since, together, mass and concentration fully specify the equilibrium
mass profile of a halo, there has been considerable interest in
understanding the origin of such correlation, as well as its
dependence on redshift and on cosmological parameters.  At given
redshift, concentration decreases monotonically with increasing halo
mass in a manner that suggests a link between the characteristic
density of a halo and the time of its assembly. Indeed, NFW showed
that the main trends in the mass-concentration relation, $c(M)$, could
be reproduced by a simple model where the characteristic density of a
halo reflects the critical density of the universe at a suitably
defined ``collapse redshift''.  The NFW model identified a halo's
collapse redshift with the time at which half the mass of the halo was
first contained in progenitors more massive than some fraction, $f$,
of its final mass. This model, however, yields acceptable results only
for surprisingly small values of $f$, of order one percent.

The NFW prescription also predicts that, at given mass, concentrations
should evolve weakly with redshift, at odds with the stronger
dependence on redshift reported in later work.  \citet{Bullock2001},
for example, argued that concentrations of halos of fixed mass should
scale linearly with expansion factor; $c\propto a$. \citet{Eke2001},
on the other hand, proposed that concentrations should depend both on
the amplitude as well as on the shape of the power spectrum. Their
predictions, however, also failed to reproduce the findings of
subsequent simulation work. It has now become clear that concentration
depends on mass and redshift in a complex fashion; for example,
although the concentration of rare, very massive halos barely evolves
with redshift, that of low mass halos evolves rapidly \citep[see,
e.g.,][]{Gao2008}.

The latter behavior is better reproduced by models that link the
concentration of a halo with its past accretion history
\citep{Wechsler2002,Zhao2003a,Lu2006}. In these models the
concentration is empirically found to trace the time when halos
transition from a period of ``fast growth'' to another where mass is
accreted more gradually. These empirical models are able to account for the
nearly constant\footnote{Recently, \citet{Prada2012} have argued for
  an ``upturn'' in the concentration of very massive halos. This has
  now been shown to be due to unrelaxed halos in transient stages of
  their evolution \citep{Ludlow2012}. The upturn disappears when only
  relaxed halos are considered.} concentration of very massive halos
(they are all still in the fast-growth phase) and for the redshift
dependence of the concentration at different halo masses.

Attempts to formalize these results into an analytic model have been
less successful. For example, as in the original NFW papers, the model
of \citet{Zhao2009} links the concentration with a time when a halo
had only assembled a surprisingly small ($\sim 4\%$) fraction of
its final mass. This casts into doubt how the model should be
interpreted, or how it should be extended to halo mass regimes or
cosmological parameters not calibrated directly by simulation.

We have recently used the Millennium Simulation series of cosmological
simulations to revisit these issues \citep{Ludlow2013}. The key result
of that work is that the {\it shape} of the mass accretion history of
a halo (hereafter MAH for short) is indistinguishable from the {\it
  shape} of its mass profile at the final time. The similarity in
shape becomes readily apparent when expressing the evolution of the
mass of the main progenitor in terms of the critical density rather
than redshift, $M(\rho_{\rm crit}(z))$, and the mass profile in terms
of enclosed mass and mean inner density, $M(\langle \rho
(<r)\rangle)$. {\it Both} shapes resemble closely the NFW profile.

This insight provides a compelling explanation for the NFW profile and
its self-similar nature, which can be traced to the mass invariance of
halo mass accretion histories \citep[see, e.g.,][]{vandenBosch2002}.
For given halo mass, its average mass accretion history may be fully described
by specifying a single ``concentration'' parameter, $c_{\rm MAH}$, and
there is a unique correspondence between this parameter and the NFW
concentration parameter, $c_{\rm NFW}$, of the halo mass profile. Once
this relation has been calibrated, concentrations may be predicted for
halos of any mass and at any redshift, and for any cosmology, provided
that accurate mass accretion histories are available.

We explore these issues here by extending our earlier $z=0$ analysis of the
Millennium Simulation series to higher redshift. We use this analysis
to calibrate the $c_{\rm MAH}$-$c_{\rm NFW}$ relation and to show how analytic
mass accretion histories may be used to predict concentration-mass
relations that are in full agreement with published results.

We begin with a brief summary of the Millennium Simulation series
(Sec.~\ref{SecSims}) and of our analysis technique
(Sec.~\ref{SecAnal}), followed by a presentation of our main results
in Sec.~\ref{SecRes}. We conclude with a brief summary of our main
conclusions in Sec.~\ref{SecConc}.

\begin{figure}
  \begin{center}
    \resizebox{8cm}{!}{\includegraphics{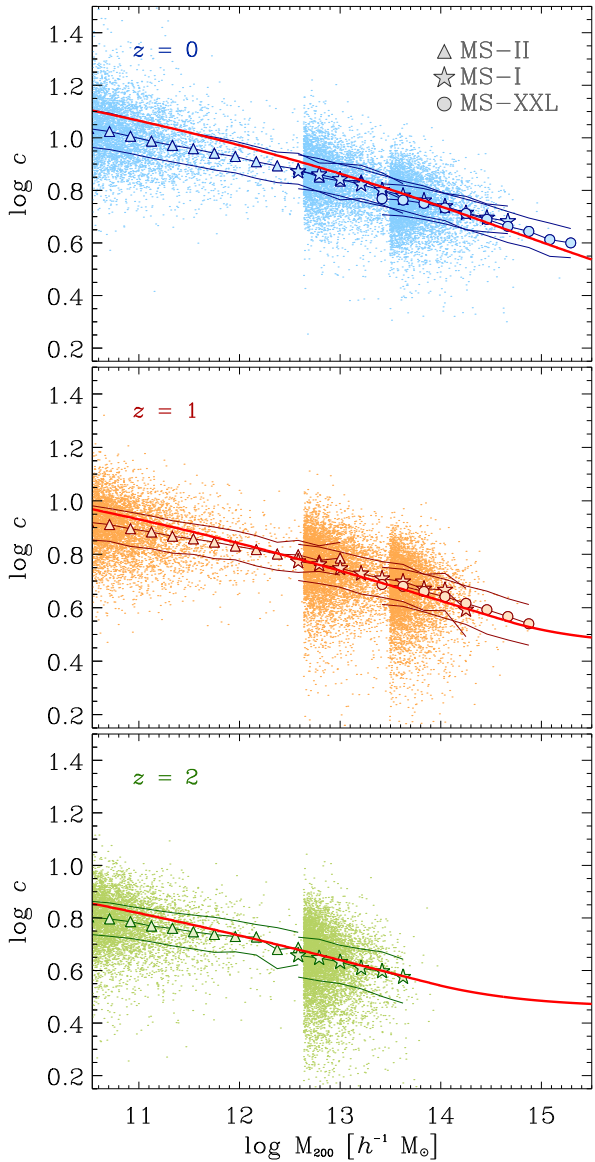}}
  \end{center}
  \caption{Mass and redshift dependence of best-fit NFW concentrations
    for all relaxed MS halos with more than $5000$ particles within
    the virial radius. From top to bottom, panels correspond to halos
    identified at $z_i=0$, $1$ and $2$, respectively. Individual halos
    are shown as colored dots; heavy filled symbols show the median
    trends for each simulation (see legend); 25th and 75th
    percentiles are shown as thin lines.  The solid red line in each
    panel shows the mass-concentration relation obtained from fitting
    NFW profiles to the mass accretion histories predicted by the
    model described in the Appendix. See Secs.~\ref{SecModMAH} and ~\ref{SecPredcMz} for more
    details.}
  \label{FigMcz}
\end{figure}

\begin{figure}
  \begin{center}
    \resizebox{8cm}{!}{\includegraphics{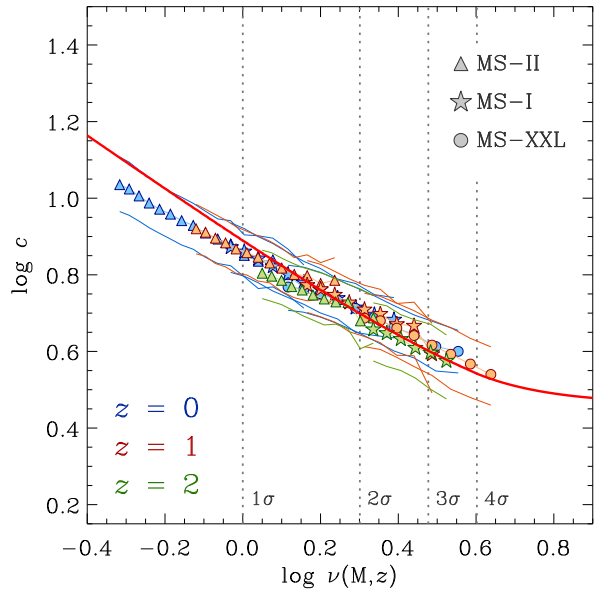}}
  \end{center}
  \caption{Halo concentrations shown as a function of the
    dimensionless mass parameter, $\nu=\delta_{\rm
      crit}(z)/\sigma(M,z)$ for all three redshifts shown in
    Fig.~\ref{FigMcz}. Heavy symbols show the same median trends as in
    Fig.~\ref{FigMcz} for each simulation, as indicated in the legend;
    thin lines show the 25th and 75th percentiles.
    The solid red line shows the $c(\nu)$ relation predicted using the
    model described in the Appendix (see
    Secs.~\ref{SecModMAH} and ~\ref{SecPredcMz} for details).}
  \label{FigCNu}
\end{figure}

\begin{center}
 \begin{table}
   \caption{Numerical properties of the Millennium and Aquarius simulations. N$_p$ is the total number of
     particles; $L_{\rm box}$ is the simulation boxsize; $\epsilon$ is the Plummer-equivalent gravitational
     force softening and $m_{\rm p}$ is the particles mass. In the case of the Aquarius runs N$_p$ is the
     number of high-resolution particles and $m_{\rm p}$ is their mass. }
   \begin{tabular}{l c c c c c c c}\hline \hline
     Simulation     & N$_p$               & $L_{\rm box}$ & $\epsilon$ & $m_{\rm p}$             & \\
                    &                     & [Mpc$/h$]     & [kpc$/h$]
                    & [M$_\odot/h$] & \\ 
\hline
     MS-XXL         &  6720$^3$           & 3000 & 10    & 6.17$\times     10^9$ & \\ 
     MS-I             &  2160$^3$           & 500  & 5     & 8.61$\times 10^8$ & \\
     MS-II          &  2160$^3$           & 100  & 1     & 6.89$\times 10^6$ & \\
     Aq-A-2         &  5.3$\times 10^8$ & -    & 0.050 & 1.00$\times 10^4$ & \\
     Aq-A-1         &  4.3$\times 10^9$ & -    & 0.015 & 1.25$\times 10^3$ & \\
\hline
   \end{tabular}
   \label{TabSimParam}
 \end{table}
\end{center}

\begin{center}
  \begin{table}
    \caption{Summary of the parameters adopted for the cosmological
      models mentioned in this paper. All models assume a flat geometry,
      so that $\Omega_{\Lambda}=1-\Omega_{\rm M}$.}
    \begin{tabular}{l c c c c c c c}\hline \hline
      Cosmology  & $\Omega_{\rm M}$ & $\Omega_{\rm bar}$ & $\sigma_8$ & $h$    & $n_s$   & \\ \hline
      Millennium &  0.250     &   0.045            &   0.90     & 0.73   &  1.0    &  \\
      WMAP 1     &  0.268     &   0.044            &   0.90     & 0.71   &  1.0    &  \\
      WMAP 3     &  0.238     &   0.042            &   0.75     & 0.73   &  0.95   &  \\
      WMAP 5     &  0.258     &   0.0441           &   0.796    & 0.719  &  0.963  &  \\
      WMAP 7     &  0.270     &   0.0469           &   0.82     & 0.70   &  0.95   &  \\
      Planck     &  0.3086    &   0.0483           &   0.8288   & 0.6777 &  0.9611 &  \\ \hline
    \end{tabular}
    \label{TabCosmoParam}
  \end{table}
\end{center}

\section{Numerical Simulations}
\label{SecSims}

Our analysis uses spherically-averaged mass profiles and mass
accretion histories of dark matter halos identified in the Millennium
Simulation suite: MS-I \citep{Springel2005a}, MS-II
\citep{Boylan-Kolchin2009}, and MS-XXL \citep{Angulo2012} (hereafter
referred to collectively as MS). Here we provide a brief description
of the simulations and their associated halo catalogs, and refer the
interested reader to those papers for further details.

\subsection{The Millennium Runs}
\label{SecMillSims}

The three MS runs adopt the same snapshot output sequence and
cosmological parameters, which were chosen to be consistent with a
WMAP-1  LCDM model: $\Omega_{\rm M}=0.25$;
$\Omega_{\Lambda}=1-\Omega_{\rm M}=0.75$; $\sigma_8=0.9$; $n_s=1$;
$h=0.73$. Here $\Omega_i$ is the contribution to the current
matter-energy density of the Universe from component $i$, expressed in
units of the critical density for closure;
$\sigma_8$ is the rms mass fluctuation in 8\Mpch \, spheres, linearly
extrapolated to $z=0$; $n_s$ is the spectral index of primordial
density fluctuations; and $h$ is the present-day Hubble expansion rate
in units of $100{ \, \rm km \, s^{-1} \, Mpc^{-1}}$.

The MS-I and MS-II evolved the dark matter density field using
N$_p=2160^3$ particles; these runs differ only in box size, $L_{\rm
  box}$, Plummer-equivalent force softening, $\epsilon$, and particle
mass, $m_{\rm p}$.  MS-XXL is the largest of the three runs both in
particle number, $N_p=6720^3$, and in box size, $L_{\rm box}=3\,
h^{-1} {\rm Gpc}$. Because of its size, however, MS-XXL particle data
was not stored for all snapshot redshifts. We list the most important
numerical parameters of these simulations in Table~\ref{TabSimParam}.

\begin{figure}
  \begin{center}
    \resizebox{8cm}{!}{\includegraphics{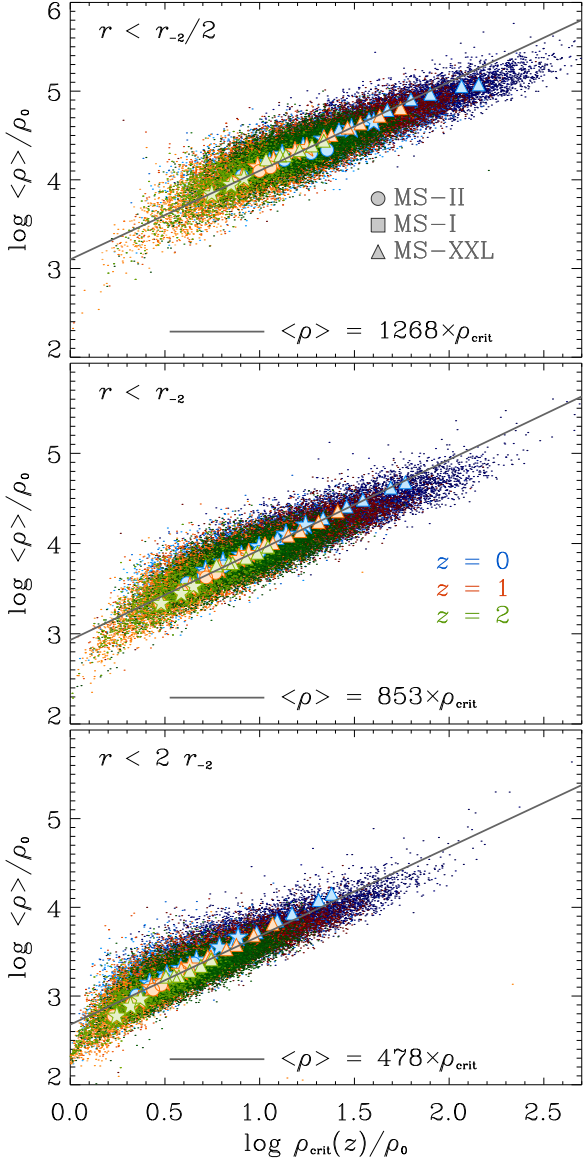}}
  \end{center}
  \caption{The mean enclosed density, $\langle \rho_{-2} \rangle$,
    within the NFW scale radius, $r_{-2}$ (middle panel), as well as
    within $r_{-2}/2$ (top panel) and $2\times r_{-2}$ (bottom panel),
    shown as a function of the critical density of the Universe at the
    time when the main progenitor first reaches the mass enclosed
    within each of those radii. Critical densities are scaled to the
    value at which a halo is identified, $\rho_0\equiv \rho_{\rm
      crit}(z_i)$. Only relaxed halos with more than $20000$ particles
    identified at redshifts $z_i=0$ (blue points), $1$ (red points)
    and $2$ (green points) are shown in order to minimize overlap.
    Note the direct proportionality between the two densities that
    holds in all panels. This implies that mass accretion histories
    can be reconstructed from the mass profiles and
    vice versa. Individual halos are shown as colored dots; large
    filled symbols show the results obtained after averaging halo mass
    profiles and accretion histories in logarithmic mass bins of width
    $\delta\log M=0.4$.}
  \label{FigRhoRho}
\end{figure}

\subsection{Dark Matter Halo Catalogs}
\label{SecHaloCatalogs}

Dark matter halos were identified in each simulation output using a
friends-of-friends (FOF) halo finder with a link-length $b=0.2$ times
the mean inter-particle separation. The subhalo finder
\textsc{subfind} \citep{Springel2001b} was then run on each FOF group
that contained at least N$_{\rm min}=20$ particles in order to
identify its self-bound substructures. \textsc{subfind} outputs a
number of useful characteristics for each FOF group and its
subgroups. We retain for our analysis only the main halo of each FOF
group (i.e., we do not consider substructure halos) and record their
virial mass, $M_{200}$, as well as the radius, $r_{\rm max}$, of the
peak circular velocity, $V_{\rm max}$. We note that $M_{200}$ refers
to the {\it total} mass within $r_{200}$, including substructure. For
a halo that follows the NFW profile, $V_{\rm max}$ and $r_{\rm max}$
are fully determined by the virial mass and concentration, and
therefore may also be used as alternative parameters to express the
halo mass profile.

Because of the dynamical nature of the formation process, DM halos
are, at best, quasi-equilibrium systems. We therefore compute three
diagnostics that can be used to flag halos that are far from
equilibrium. Including halos in such transient states in the analysis
can lead to biased estimates of their mean structural properties,
especially at large halo masses, where a majority of the systems may
be out of equilibrium
\citep[e.g.,][]{Neto2007,Maccio2008,Ludlow2012}. The ``relaxation''
diagnostics are: (i)  the substructure mass fraction, $f_{\rm sub}=M_{\rm
  sub}(<r_{200})/M_{200}$, (ii) the center of mass offset from the
potential minimum, $d_{\rm off}=|\mathbf{r}_p-\mathbf{r}_{\rm
  CM}|/r_{200}$, and (iii) the virial ratio of kinetic to potential
energies, $2\, K/|U|$. In the analysis that follows we shall only
consider halos that simultaneously satisfy the following three
conditions: i) $f_{\rm sub}<0.1$, ii) $d_{\rm off}<0.07$, and iii)
$2\, K/|U|<1.35$.

The relative importance of unrelaxed halos increases with halo mass
and with redshift, reflecting the more recent assembly of rarer, more
massive structures. For example, fewer than $50\%$ of halos with virial
mass in excess of $\sim 10^{14}\, h^{-1} \, M_\odot$ pass our three
relaxation criteria at $z=0$.  By $z=1$, this fraction decreases to
$\sim 20\%$. Similarly, roughly $80\%$ of halos with present-day
virial mass of order $10^{12}\, h^{-1} \, M_{\odot}$, meet
our three relaxation criteria.  For the same mass scale, the fraction
decreases to $\sim 52\%$ at $z=1$; by $z=2$ only about one third of
these systems are deemed relaxed. We exclude unrelaxed halos from our
analysis since they are only poorly approximated by simple fitting
formulae like the NFW profile, and because their transient state means
that their fit parameters change quickly, hampering interpretation.

\begin{figure*}
  \begin{center}
    \resizebox{16cm}{!}{\includegraphics{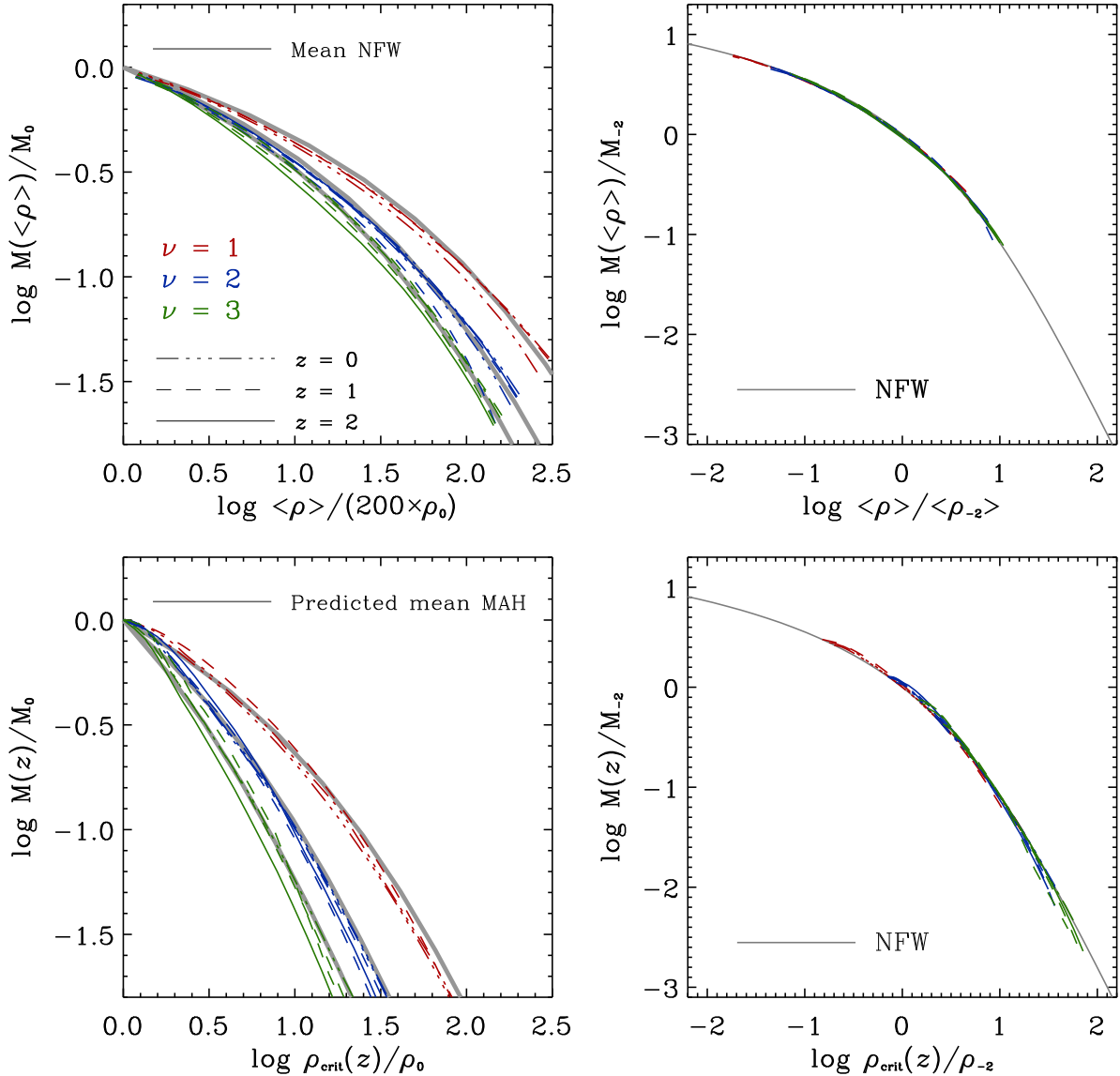}}
  \end{center}
  \caption{Average mass profiles and accretion histories for halos
    with dimensionless mass parameter values equal to $\nu=1$ (red),
    $2$ (blue), and $3$ (green). Different line styles correspond to
    halos identified at different redshifts. Top panels show enclosed
    mass versus mean inner density profiles; bottom panels show the
    growth of the mass of the main progenitors as a function of
    redshift, expressed in units of the critical density. All masses
    are scaled to the final mass of the halo at the time of
    identification. Densities in the bottom left panel are scaled to
    the critical density at the identification redshift,
    $\rho_0=\rho_{\rm crit} (z_i)$. Densities in the top left panel
    are scaled to the mean enclosed density at the virial radius,
    $200\times \rho_0$. The top right-hand side panel shows the same
    profiles as on the left, but scaled to the mass and density within
    the scale radius of the best-fitting NFW profile. The bottom right
    shows the same curves as on the left, scaled, as in the top-right
    panel, to the characteristic mass and density of the NFW best
    fit. Note that, independent of halo mass and of identification
    redshift, both the mass profiles and accretion histories follow
    closely the same NFW shape. Note as well that the accretion
    histories of all halos with the same value of $\nu$ are similar,
    regardless of redshift.}
  \label{FigMrhoMz}
\end{figure*}

\section{Analysis}
\label{SecAnal}

\subsection{Mass Profiles and Accretion Histories}
\label{SecMassProfiles}

We have computed spherically-averaged density profiles and mass
accretion histories for relaxed halos identified at three different
redshifts: $z_i=0$, $1$, and $2$. Density profiles are computed in
$32$ radial bins that span the range $-2.5 < \log r/r_{200} < 0$ in
equally spaced steps in $\log r$. Within each radial bin we also
compute the total enclosed mass, $M(r)$ (including substructure), and the mean inner density,
$\langle\rho\rangle(r)=M(r)/(4/3)\pi \, r^3$.  In order to obtain
robust estimates of the structural parameters of the halo mass
profiles, we restrict our analysis to systems that contain at least
5000 particles within their virial radius, $r_{200}$. Even with these
strict limits there are many thousands of halos in our sample at each
redshift: at $z_i=0$ about 1.1 million halos are included in our
sample. At $z_i=1$ and $2$, the numbers go down to roughly $300000$
and $32000$, respectively. (MS-XXL is excluded from our $z_i=2$
analysis because particle data is unavailable for this redshift.)

We compute the mass accretion history (MAH) of each halo identified at
$z_i$ by tracking the virial mass of its main progenitor
through all previous simulation outputs. Although other operational
definitions of assembly histories exist \citep[see,
e.g.][]{Wang2011,Giocoli2012}, ours is simple to compute both in
simulations as well as in semi-analytic models for structure
formation, such as those based on excursion-set theory
\citep{Kauffmann1993,Lacey1993,EisensteinLoeb1996,Nusser1999,Somerville1999a}.

As mentioned in Sec.~\ref{SecIntro}, although the MAH is most often
written as a function of redshift or expansion factor, we will instead
express it in terms of the critical density of the Universe; i.e., as
$M(\rho_{\rm crit}(z))$.  This facilitates a comparison between the
{\em shape} of the mass accretion history and that of mass profile,
which we write in terms of the mean enclosed density, $M(\langle\rho \rangle)$.

\subsection{NFW Profiles and Fitting Techniques}
\label{SecNFWProfiles}

We estimate halo structural parameters by fitting eq.~(\ref{EqNFW}) to
the mass profiles of the full sample of equilibrium halos 
described above. The NFW profile has two free parameters, which are 
simultaneously adjusted in order to minimize the figure-of-merit function
\begin{equation}
  \psi^2 = \frac{1}{N_{\rm bin}} \sum_{i=1}^{N_{\rm bin}} [\ln \rho_i - \ln \rho_{\rm NFW}(\delta_c;r_{s})]^2.
  \label{EqFoM_rho}
\end{equation}
In order to avoid biases that may be introduced by substructures in
the outskirts of the halo we restrict our fits to the radial range
$r_{\rm min} < r < 0.6 \, r_{200}$, where $r_{\rm min}$ is chosen to
be either the ``convergence'' radius, $r_{\rm conv}$\footnote{The
  ``convergence radius'' is defined here as in \citet{Power2003}.}, or
$0.05\times r_{200}$, whichever is smaller \citep[for a more detailed
discussion of the implications of limiting the outer fit radius please
see][]{Ludlow2010}.  In general, NFW fits to the spherically averaged
$\rho(r)$ profiles are good: at $z=0$, for example, $\psi_{\rm min}$
has a median value of $0.083^{+0.019}_{-0.016}$, where the quoted
range indicates the upper and lower quartiles of the
distribution. Similarly good fits are obtained at redshifts $z_i=1$
and $2$. The best-fit NFW profiles yield estimates of the halo
structural parameters $\delta_c$ and $r_{s}$ for each halo in our
sample, which we use in turn to estimate $c_{\rm NFW}$, the concentration parameter.

As discussed above, NFW profiles can also be fit to halo mass
accretion histories in order to estimate their ``concentration''
parameter \citep{Ludlow2013}. We do so by fitting the MAH, written as
$M(\rho_{\rm crit})$, with the NFW profile expressed in terms of the
enclosed density:
\begin{equation}
  \langle\rho\rangle (r) = \frac{3 M(<r)}{4 \, \pi\, r^3} = \frac{200}{x^3}\frac{g(c\, x)}{g(c)}\rho_{\rm crit},
  \label{EqNFW_enc}
\end{equation}
where $x\equiv r/r_{200}$ and $g(y)=\ln (1+y)-y/(1+y)$. Since we are
interested in estimating a single parameter, the concentration, we
first normalize the MAH to the halo virial mass, $M_0=M(z_i)$, 
and critical density, $\rho_0=\rho_{\rm crit}(z_i)$, at the redshift of 
interest, and then determine the value of the remaining parameter, 
$c_{\rm MAH}$, by minimizing the rms deviation between eq.~(\ref{EqNFW_enc})
and the MAH. 

\begin{figure}
  \begin{center}
    \resizebox{8cm}{!}{\includegraphics{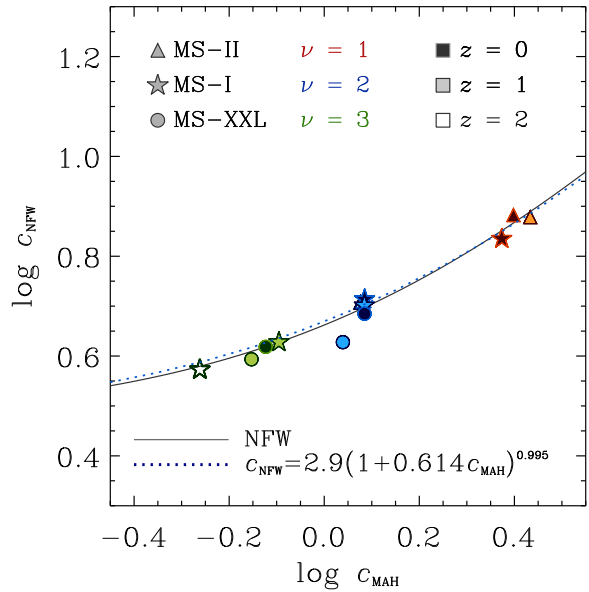}}
  \end{center}
  \caption{Relation between the concentration parameters obtained from
    NFW fits to the mass profiles ($c_{\rm NFW}$) and accretion
    histories ($c_{\rm MAH}$) shown in Figure~\ref{FigMrhoMz}. Individual points are colored according to peak
    height, $\nu$; filled or open symbols indicate the redshift and
    simulation of each halo sample, as indicated in the legend. The
    solid curve shows the relation expected given the
    correlations shown in the middle panel of Fig.~\ref{FigRhoRho};
    the dotted line shows a useful approximation to this curve:
    $c_{\rm NFW} = 2.9\times (1+0.614\, c_{\rm MAH})^{0.995}$.}
  \label{Figcc}
\end{figure}

\section{Results}
\label{SecRes}

\subsection{The mass-concentration-redshift relation}
\label{SecMcz}

\begin{figure*}
  \begin{center}
    \resizebox{16cm}{!}{\includegraphics{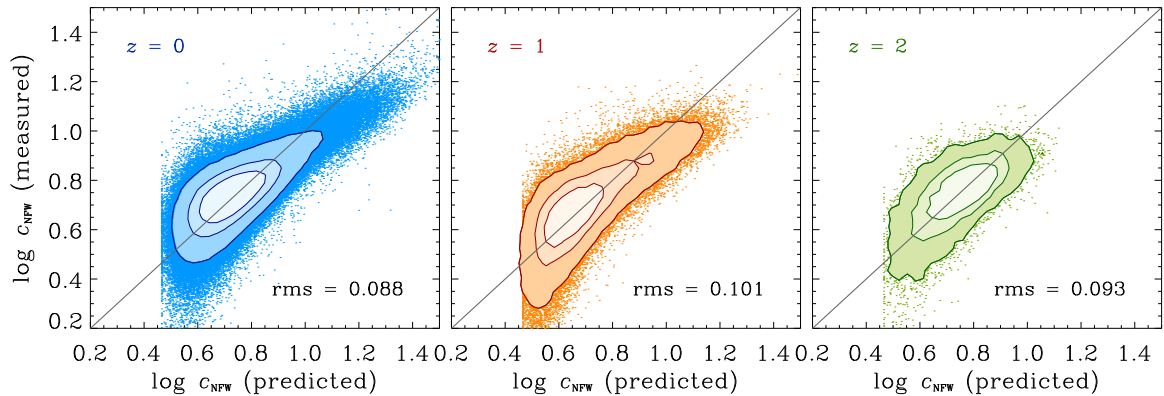}}
  \end{center}
  \caption{A halo-to-halo comparison of the measured concentrations and those predicted by 
    fitting NFW profiles to the mass accretion histories and using the
    relation shown in Fig.~\ref{Figcc}. Contours enclose 50, 75 and 95 percent
    of the halos, with the remaining 5 percent shown as individual points. The rms scatter
    in the ratio of measured-to-predicted concentration is given in each panel.}
  \label{Figlgclgc}
\end{figure*}

The mass-concentration relation, $c(M)$, is shown in Fig.~\ref{FigMcz}
for equilibrium halos identified at $z_i=0$, $1$, and $2$. 
The heavy symbols trace the median trend for each halo sample, and the 
thin lines delineate the 25th and 75th percentile range. Note 
the excellent agreement between different simulations in the mass intervals 
where they overlap, a clear demonstration that our results are not
compromised by numerical artifact. We find, in agreement with earlier
work, that the concentration is a monotonic
\footnote{The ``plateau'' in concentration at large halo masses
  reported in earlier work is not apparent here because we are only
  considering relaxed halos; including unrelaxed massive halos raises
  the median concentration. See \citet{Ludlow2012} for further
  details.}
but weak function of mass, varying by only a factor of $\sim 4$ over
the $6$ decades in mass resolved by the simulations at
$z=0$. Concentrations decrease systematically with increasing redshift
over the mass range resolved here, although the magnitude of the
decrease is mass-dependent: from $z=0$ to $z=2$ concentrations
decrease by about a factor of $1.6$ for $10^{11}\, h^{-1} \, M_\odot$
halos but less than $50\%$ for $10^{14}\, h^{-1} \, M_\odot$ halos.

This seemingly complex mass-dependent evolution actually reflects a
simpler dependence, which becomes apparent when expressing
concentrations in terms of the dimensionless ``peak height'' mass
parameter,
\begin{equation}
\nu(M,z)=\delta_{\rm crit}(z)/\sigma(M,z),
\label{EqNu}
\end{equation}
defined as the ratio between the critical overdensity for collapse at
redshift $z$ and the linear rms fluctuation at $z$ in spheres of mass
$M$. The larger $\nu$ the rarer the halo, and the more massive it is
in relation to the characteristic clustering mass, $M_\star$, defined
by $\nu(M_\star)=1$.

Fig.~\ref{FigCNu} shows that concentrations depend primarily on $\nu$,
over the entire range of masses and redshifts analyzed here \citep[for
a similar result, see also][and references therein]{Prada2012}.  This
result is consistent with the rescaling procedure outlined by
\citet{Angulo2010} to match the outcome of simulations adopting
slightly different power spectra and cosmological parameters.  It also
offers an important clue that any successful analytical model of the
mass-concentration-redshift relation should reproduce\footnote{Note that halo collapse times (and therefore
      concentrations) may also depend on the expansion history of the
      Universe \citep{Dolag2004}. A residual, but weak, dependence on
      the shape of the linear growth factor might therefore become
      evident for expansion histories that differ significantly from
      $\Lambda$CDM.}.  The thick red lines in Figs.~\ref{FigMcz} and
~\ref{FigCNu} present the predictions of one such model, which we
discuss in detail in Sec.  ~\ref{SecPredcMz} below.

\subsection{MAH and Mass Profiles}
\label{SecMAHMprof}

As discussed by \citet{Ludlow2013}, the shape of the mass
profiles of MS halos identified at $z=0$ is intimately related to
their accretion histories. We show in Fig.~\ref{FigRhoRho} that this
result also applies to halos identified at higher redshift. The middle
panel of this figure plots the mean density enclosed within the NFW
scale radius, $\langle \rho_{-2}\rangle=M_{-2}/(4/3)\pi r_{-2}^3$,
versus $\rho_{\rm crit}(z_{-2})$, the critical density of the Universe
at the time when the mass of the main progenitor halo equals
$M_{-2}$. Results for all three identification redshifts are included,
after scaling densities to the critical density at each $z_i$, $\rho_0
\equiv \rho_{\rm crit}(z_i)$.

Note that the two densities scale linearly, and that there is no
difference between halos selected at different $z_i$. A similar result
is found when repeating the exercise within a radius equal to one half
or twice the scale radius (top and bottom panels of
Fig.~\ref{FigRhoRho}, respectively). This confirms that mass accretion
histories can be reconstructed from the mass profile, and vice versa,
for halos selected at any redshift. 

Further, as Fig.~\ref{FigMrhoMz} makes clear, mass accretion histories
and mass profiles have the {\it same} shape on average, which can be
closely approximated by the NFW profile. In this figure, the
similarity becomes apparent when comparing MAHs expressed as
$M(\rho_{\rm crit})$ with mass profiles expressed as $M(\langle \rho
\rangle)$. Each of these profiles, shown for different values of $\nu$
in the left panels, can be scaled to the characteristic values of
their best NFW fits. This is shown in the right-hand panels, which
puts in evidence the remarkable similarity between MAH and mass
profile shapes.

\begin{figure}
  \begin{center}
    \resizebox{8cm}{!}{\includegraphics{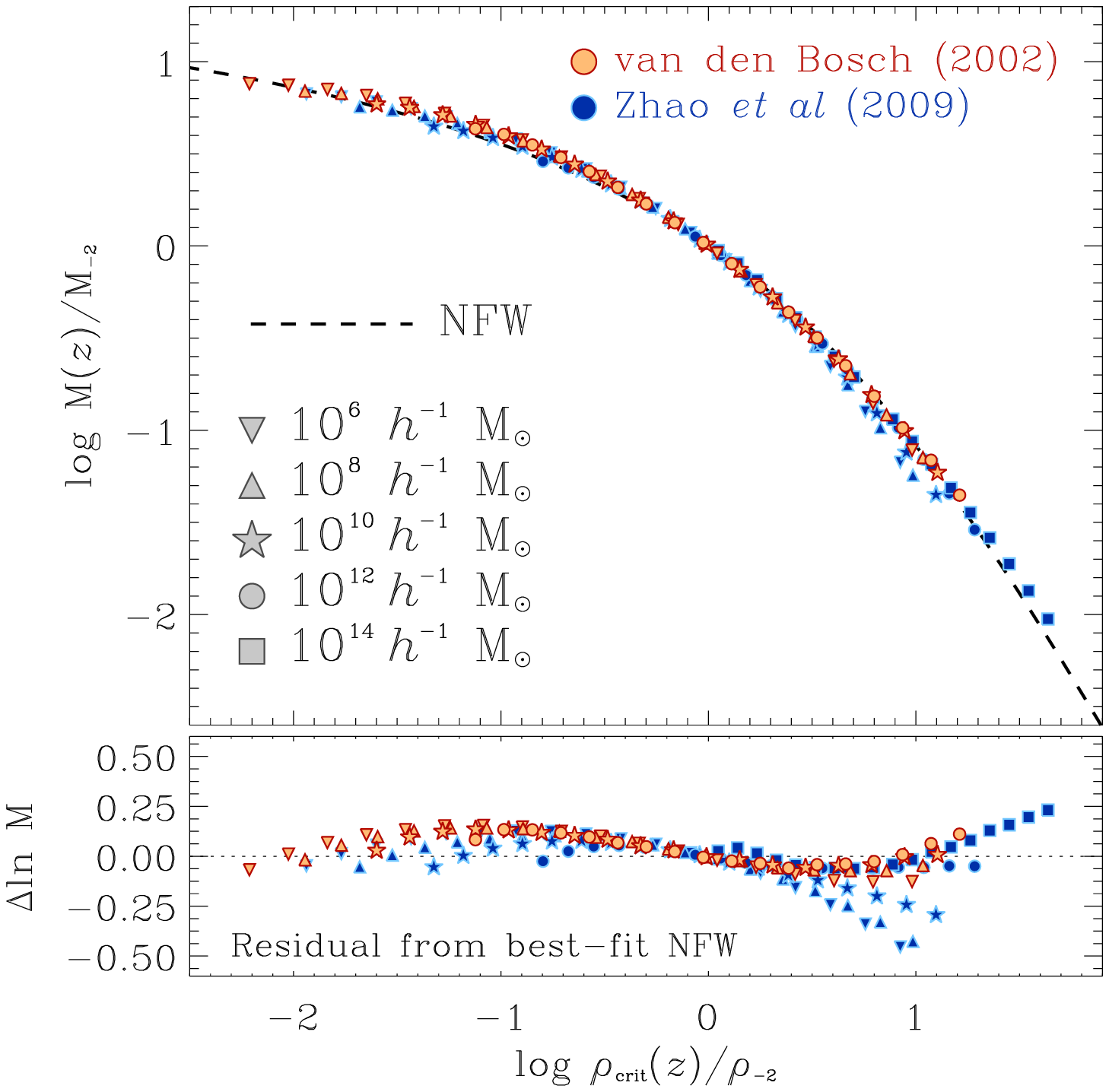}}
  \end{center}
  \caption{Mass accretion histories for dark matter halos predicted by
    the models of \citet{Zhao2009} (blue symbols) and
    \citet{vandenBosch2002} (orange symbols), scaled to their best-fit
    NFW parameters: $M_{-2}$ and $\rho_{\rm crit} (z_{-2})$. Assembly
    histories are shown for halos spanning eight orders of magnitude
    in present-day mass, ranging from $10^6 \ h^{-1}$ M$_{\odot}$ to
    $10^{14} \ h^{-1}$ M$_{\odot}$. The dashed line shows the NFW
    profile, which is fixed in these units. Residuals from best-fit
    NFW profiles are shown in the bottom panel.}
  \label{FigModelMAH}
\end{figure}

Fig.~\ref{FigMrhoMz} also shows that average accretion histories are
solely a function of the dimensionless mass parameter $\nu$. Indeed,
regardless of the identification redshift, the scaled mass accretion
histories of halos with the same value of $\nu$ are essentially
indistinguishable from each other. This explains why concentrations
depend only on $\nu$, as shown in Fig.~\ref{FigCNu}: accretion
histories fully determine the final mass profile of a halo and, since
MAHs depend only on $\nu$, so do concentrations.

The results discussed above imply that one may use halo mass profiles
to predict, on average, their MAH. Since both are well approximated by
the NFW profile, all that is needed, for given halo mass, is to
calibrate the relation between the MAH ``concentration'' and that of
the mass profile. This may be derived analytically from the
proportionality constant between $\langle \rho_{-2} \rangle$ and
$\rho_{\rm crit}(z_{-2})$ shown in the middle panel of
Fig.~\ref{FigRhoRho}. The result is shown by the solid line in
Fig.~\ref{Figcc}, together with a simple approximation,
\begin{equation}
c_{\rm NFW} = 2.9\, (1+0.614\, c_{\rm MAH})^{0.995},
\label{Eqcc}
\end{equation}
that proves accurate over the range of concentrations probed by our
simulations.  

\begin{figure*}
  \begin{center}
    \resizebox{16cm}{!}{\includegraphics{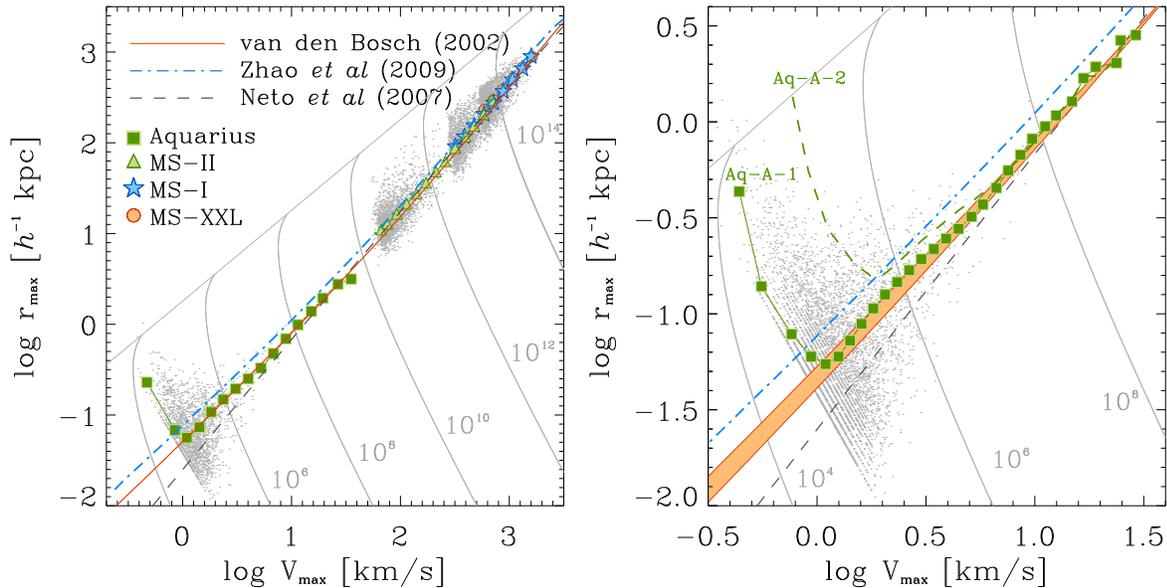}}
  \end{center}
  \caption{{\em Left panel:} Relation between $r_{\rm max}$ and
    $V_{\rm max}$ at $z=0$ for isolated dark matter halos in the
    Millennium Simulations and in the Aq-A simulation of the Aquarius
    Project \citep{Springel2008b}. Individual halos are shown by the
    grey points; grey curves indicate halos of constant virial mass,
    labeled in units of $h^{-1}\, M_\odot$. Heavy symbols trace the
    median $r_{\rm max}$ and $V_{\rm max}$, computed in equally spaced
    logarithmic bins of halo mass. The upturn in the Aquarius relation
    at low masses indicates the onset of artifacts induced by limited
    numerical resolution. The gray dashed line indicates the $r_{\rm
      max}$-$V_{\rm max}$ relation corresponding to the power-law
    $c(M)$ relation derived by \citet{Neto2007} from halos in the MS-I
    simulation. The solid orange line shows to the relation predicted
    by NFW-fitting the mass accretion histories of
    \citet{vandenBosch2002}; the dot-dashed blue line corresponds to
    MAHs predicted by \citet{Zhao2009}. {\em Right panel:} A zoom-in
    of the left panel showing only isolated halos in the Aquarius
    simulation. These are defined as systems that, at $z=0$, lie at
    least three virial radii away from the main Aquarius halo.
    Symbols and lines are as in the left.  The orange shaded area
    shows how the \citet{vandenBosch2002} relation changes when
    varying the MAH range fitted; from $10^{-3}<M/M_0<0.8$ (top) to
    $0.1<M/M_0<0.8$ (bottom).}
  \label{FigVmaxRmax}
\end{figure*}

A couple of points are worth remarking about this relation. The first
is that it predicts a minimum concentration ($c_{\rm NFW} \sim 2.9$) 
for halos that form relatively recently and whose MAH, consequently, is
best described with a very low value of $c_{\rm MAH}$. 
This is broadly consistent with the idea, first proposed by NFW, that
there should be little difference in the concentration of very massive
systems ($M\gg M_\star$), if concentration in any way reflects the
formation time. Indeed, since all of these halos must be collapsing
just before the time at which they are identified, they share a common
assembly time and should therefore have similar
concentrations\footnote{Very massive ($\nu\simgt 3$) halos actually
  deviate slightly but systematically from the above trends; see, for
  example, how halos with the smallest values of $\langle \rho_{-2}
  \rangle$ fall below the fitted relation in
  Fig.~\ref{FigRhoRho}. Indeed, their MAH shapes differ slightly, but
  systematically, from NFW \citep{Ludlow2013}. As discussed by
  \citet{Gao2008}, the mass profiles of these systems also deviate
  from NFW and are better described by Einasto profiles with large
  values of the Einasto parameter $\alpha$.}.

The second point to note about eq.~(\ref{Eqcc}) is that the $c_{\rm
  NFW}$ dependence on $c_{\rm MAH}$ is quite shallow, implying that
even large changes in MAH lead to modest changes in mass profiles. We
show one application of this relation in the bottom-left panel of
Fig~\ref{FigMrhoMz}. The thick grey lines in this panel show the {\it
  predicted} MAHs corresponding to the mass profiles shown in the
upper-left panel, using eq.~(\ref{Eqcc}). The predictions match the
actual MAHs measured in the simulations quite well.

The procedure can also be reversed, so that accretion histories may be
used to predict mass profile concentrations. We show this in
Fig.~\ref{Figlgclgc}, where we compare, for each identification
redshift, the concentrations measured in the simulations with
estimates obtained by fitting {\it individual} halo accretion
histories and then using eq.~(\ref{Eqcc}) to predict
$c_{\rm NFW}$. As Fig.~\ref{Figlgclgc} makes clear, the predicted
concentrations for relaxed halos are in good agreement with 
measured ones; the rms between predicted and measured values is only of order 
$\sim 25\%$. This is remarkable, given the fact that (i) individual MAH are often
complex and at times not even monotonic, with several local maxima caused
by distinct merger events, and (ii) that our model attempts to describe them
with a {\it single} parameter.

\subsection{Model Mass Accretion Histories}
\label{SecModMAH}

The results of the previous subsection imply that the
concentration-mass-redshift relation can be predicted analytically
provided that accurate mass accretion histories are available. In the
recent past, this topic has received considerable attention, which has
led to the development of sophisticated algorithms able to compute
accretion histories that agree very well with the results of
cosmological N-body simulations, for arbitrary halo masses, redshift,
power spectra, and cosmological parameters. Analytic models based on
extensions of the \citet{Press1974} formalism, for example, have been
used extensively to predict halo formation times, progenitor mass
distributions, halo merger rates, and other statistics
\citep{Bond1991,Bower1991,Lacey1993,Kauffmann1993,Somerville1999a,ShethLemson1999,Parkinson2008}.
Other methods follow the hierarchical build-up of dark matter halos
using Lagrangian perturbation theory
\citep[e.g.][]{Taffoni2002,Monaco2013}, or by calibrating empirical
relations against the results of N-body simulations \citep{Zhao2009}.

For the purpose of estimating concentrations, we only need to specify
the rate at which halos increase their mass with time, and therefore
approximate methods that yield the ensemble average $\langle M(z)/M_0
\rangle$ for halos of a given mass $M_0$ would suffice. Because
accretion histories are primarily a function of the dimensionless peak
height, $\nu$, MAHs that describe halos at $z=0$ can easily be rescaled
to describe the growth of halos identified at other redshifts. We are
mainly interested in methods that reproduce, to good
approximation, our finding that $M(\rho_{\rm crit}(z))$ follows
approximately the NFW profile. Although the literature is extensive,
we focus here on two models that satisfy this criterion
\citep{vandenBosch2002,Zhao2009}, and refer the interested reader to
those papers for a more comprehensive list of references.

\begin{figure}
  \begin{center}
    \resizebox{8cm}{!}{\includegraphics{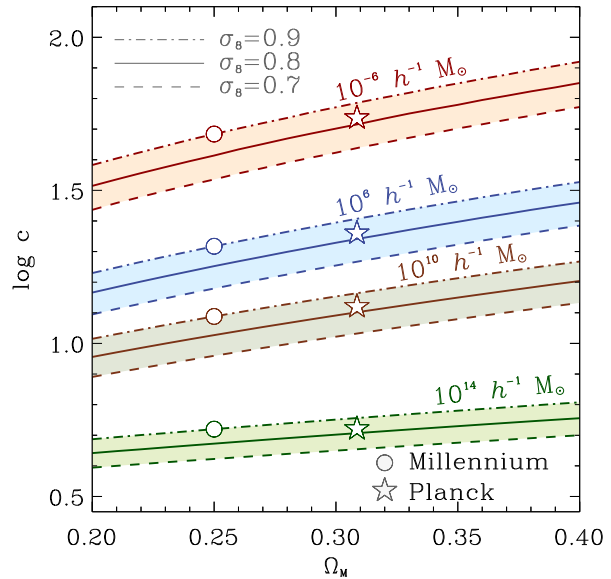}}
  \end{center}
  \caption{Concentration as a function of the matter density
    parameter, $\Omega_{\rm M}$, and the normalization of the matter
    power spectrum, $\sigma_8$, for halos of four different masses:
    $10^{-6}$ (red), $10^6$ (blue), $10^{10}$ (brown) and $10^{14} \, h^{-1} \,
    M_{\odot}$ (green). Different line-styles correspond to different
    values of $\sigma_8$, as indicated in the legend.  We assume for
    all models a flat geometry, $\Omega_\Lambda=1-\Omega_{\rm M}$. The
    remaining parameters assume values consistent with the latest
    Planck cosmology (see Table~\ref{TabCosmoParam}). Heavy symbols
    highlight the concentration values corresponding to the Millennium
    and Planck cosmologies, as indicated in the legend.  }
  \label{FigcOmegaSigma8}
\end{figure}

The \citet{vandenBosch2002} model (hereafter vdB, for short) is
largely based on the extended Press-Schechter theory, whereas the
\citet{Zhao2009} model has been constructed empirically from
cosmological N-body simulations. As shown in Fig.~\ref{FigModelMAH},
both predict NFW-like mass accretion history shapes. This figure shows
predicted MAHs for halos spanning eight orders of magnitude in mass,
after scaling them by their best fitting NFW parameters. The
deviations from the NFW shape are quite small, typically less than
$10$-$20\%$ over two decades in $M(z)$. The predictions of the two
models are essentially indistinguishable for halos above $\sim 10^8\,
M_\odot$, although the \citet{Zhao2009} model appears to deviate more
strongly from NFW at smaller masses.

These deviations imply small differences in the predicted
concentrations of low mass halos. Unfortunately, the differences
become appreciable only for masses smaller than those resolved by the
Millennium Simulations, where the smallest halos for which we can
reliably measure concentrations are of order $10^{10}\, M_\odot$. We
therefore resort to higher resolution simulations of smaller volumes,
such as those of the Aquarius Project \citep{Springel2008b}. These
simulations resolve volumes surrounding Milky Way-sized halos, and
allow us to probe the mass-concentration relation of halos with masses
as small as $10^5\, M_\odot$ or so.

We show the result of this exercise in Fig.~\ref{FigVmaxRmax}, where
we plot the $r_{\rm max}$ vs $V_{\rm max}$ for all halos in our
sample. (As discussed in Sec.~\ref{SecHaloCatalogs}, these parameters
provide an alternative characterization of the mass and concentration
of a halo.)  We also plot all {\it isolated} halos
in the high-resolution regions of the Aquarius simulations. These are
defined to be the main halos of all FOF groups found at least three
virial radii away from the center of the main Aquarius halo. This is
done in order to ensure that all subhalos ``associated'' with the main
halo are excluded from the analysis \citep{Ludlow2009}, since their
concentrations and masses were likely modified by the tidal field of
the main object.

The right-hand panel of Fig.~\ref{FigVmaxRmax} zooms-in on the
low-mass halo regime resolved only by the Aquarius runs. We use the
highest-resolution Aq-A-1 run in order to extend as much as possible
the dynamic range of the comparison. All isolated high-resolution
halos in the Aq-A volume are shown with grey points, together with
their median values after grouping them by virial mass in bins of
equal logarithmic width (solid squares). Grey lines show the loci of
halos of constant virial mass. A dot-dashed curve shows the
predictions of the \citet{Zhao2009} model, a solid curve corresponds
to the \citet{vandenBosch2002} model, while a dashed line in black shows the
simple power-law mass-concentration relation fitted by
\citet{Neto2007} to dynamically relaxed MS-I halos.

Fig.~\ref{FigVmaxRmax} makes clear that all of these models reproduce
extremely well the $r_{\rm max}$-$V_{\rm max}$ relation over roughly
ten decades in mass, although the \citet{vandenBosch2002} model seems
to outperform the others at very low masses. At $\sim 10^5 \,
M_\odot$, the Neto et al. power-law predicts halos that are more
concentrated\footnote{We note that \citet{Springel2008b} report good
  agreement between the Neto et al. power law and the Aquarius
  results. The (small) differences we remark upon here are due to the
  fact that we enforce a stricter ``isolation'' criterion and that we
  bin halos {\it by mass} rather than by $V_{\rm max}$ before
  computing their median $r_{\rm max}$.} than measured in the
simulation, whereas the Zhao et al. model deviates in the opposite
sense. Concentrations derived from accretion histories computed using
the vdB model, on the other hand, seem to predict the median
concentration as a function of mass accurately over the whole range in
halo mass resolved by these simulations. We shall therefore, in what
follows, adopt the vdB model in order to compute analytic estimates of
the concentration that can be compared with the results of
simulations. A description of the procedure may be found in the Appendix.

\subsection{Predicted mass-concentration-redshift relation}
\label{SecPredcMz}

Once a model for generating mass accretion histories has been adopted,
it is straightforward to fit them with an NFW profile in order to predict
concentrations for LCDM halos of arbitrary mass and at any
redshift. The thick solid (red) lines in Fig.~\ref{FigMcz} shows the
$c(M)$ relation at $z=0$, $1$, and $2$ predicted by the vdB model,
compared with the MS halo data. Together with Fig.~\ref{FigVmaxRmax},
Fig.~\ref{FigMcz} demonstrates that the procedure works remarkably well
at all $z$ and at all masses, especially considering the simplicity of
the model, which is based on fitting a single parameter to ensemble
average accretion histories.

We can use the same model to explore the dependence of $c(M,z)$
on cosmological parameters. This is shown in
Fig.~\ref{FigcOmegaSigma8}, where we show how the median
concentrations of $10^6$, $10^{10}$ and $10^{14} \, h^{-1}\, M_\odot$
halos at $z=0$ depend on the matter-density parameter, $\Omega_{\rm
  M}$, and the fluctuation amplitude $\sigma_8$, for a Universe with
flat geometry (i.e., $\Omega_{\Lambda}=1-\Omega_{\rm M}$; the other
cosmological parameters are assumed to take the Planck cosmology
values, see Table~\ref{TabCosmoParam}). This figure makes
clear that the changes in concentration induced by varying the
cosmological parameters is mass dependent, affecting more strongly low
mass halos than massive systems. 

The symbols in Fig.~\ref{FigcOmegaSigma8} show the predicted
concentrations for halos in the Millennium Simulation cosmology and
for the cosmological parameters favoured by latest analysis of the
Planck satellite data \citep{Planck2013}, respectively. At fixed mass,
halo concentrations increase with both $\Omega_M$ and $\sigma_8$ and,
as a result, concentrations derived from Millennium Simulation halos
at $z=0$ are actually very similar to those expected for the Planck
cosmology. At $10^{10}$ and $10^{12} \, h^{-1}\, M_\odot$, for
example, MS concentrations should be corrected upward by only $\sim \,
7\%$ and 5\%, respectively. For massive halos, with $M_0\sim 10^{14}
\, h^{-1}\, M_\odot$, both cosmological models predict $c\sim 5.3$.

\begin{figure*}
  \begin{center}
    \resizebox{15cm}{!}{\includegraphics{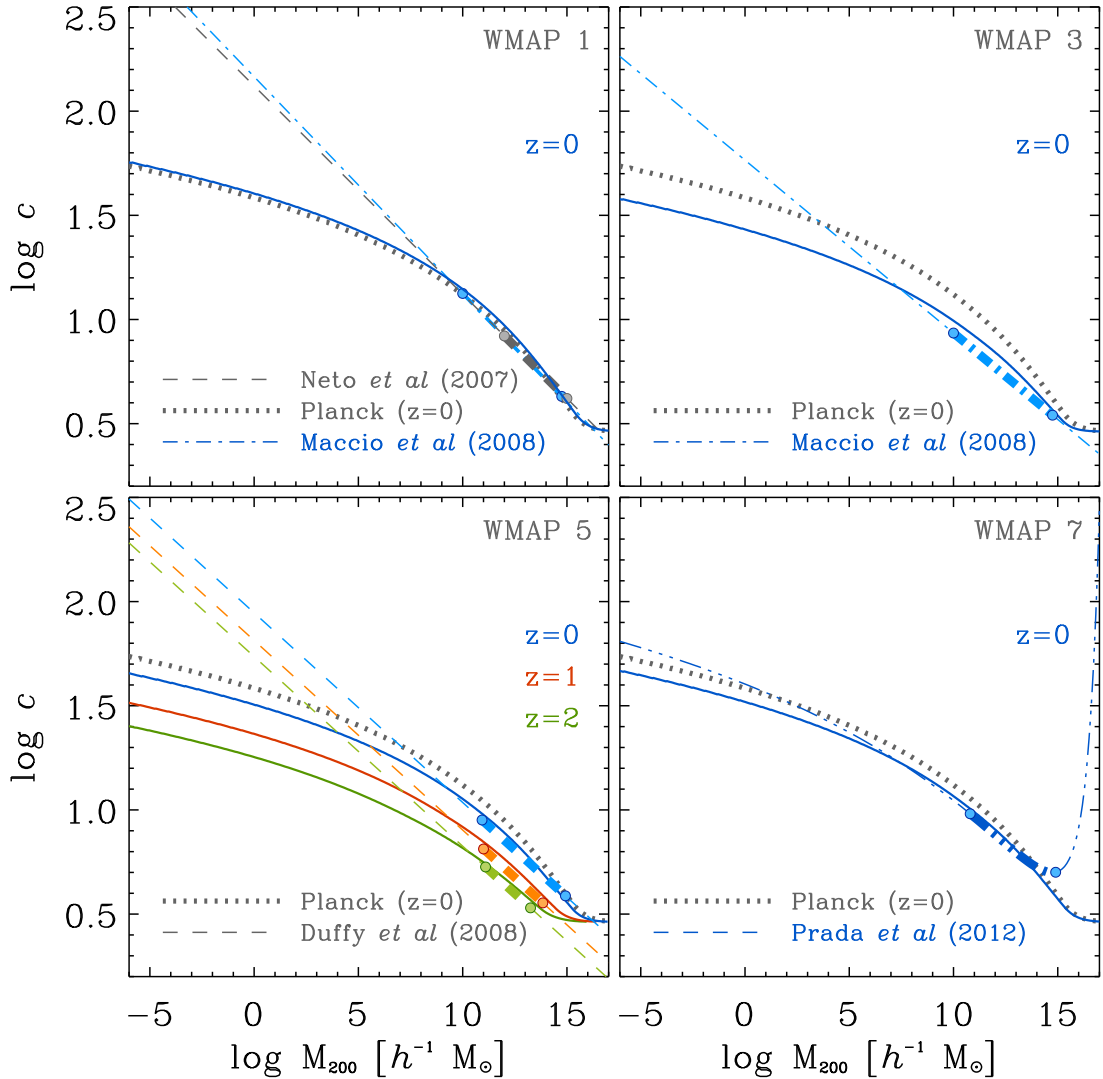}}
  \end{center}
  \caption{The concentration-mass relation from earlier simulation
    work that assumed different LCDM cosmological parameters. The
    top-left panel shows the results of two simulations at $z=0$; one
    assuming the WMAP1 cosmological parameters, from
    \citet{Maccio2008}, and the MS cosmology, from
    \citet{Neto2007}. The top-right panel shows the $c(M)$ relation
    for the WMAP3 cosmology from \citet{Maccio2008}. The bottom-left
    panel shows results for the WMAP5 cosmology from
    \citet{Duffy2008}, and the bottom-right panel corresponds to the
    WMAP7 results from \citet{Prada2012}. (We note that the
    \citet{Prada2012} model for $c(M,z)$ predicts a different redshift
    dependence that ours; we show here only results at $z=0$.)  The
    grey dotted line in each panel shows, for comparison, the
    predicted relation for the Planck cosmology.  The values of the
    cosmological parameters used for each are listed in
    Table~\ref{TabCosmoParam}. In each panel, the dashed and
    dot-dashed lines indicate the fits proposed by these authors, with
    a thick line type indicating the halo mass range actually resolved
    in each simulation. When available, we have used the results
    quoted for relaxed halos by each author. Solid lines indicate the
    predictions of our model. Note that they describe very well the
    results of earlier work over their resolved halo mass range. Our
    model, however, predicts systematic deviations at the low-mass end
    compared with simple extrapolations of power-laws fits calibrated
    at higher masses. }
  \label{FigMczComp}
\end{figure*}

The dependence of concentration on cosmology shown in
Fig.~\ref{FigcOmegaSigma8} agrees well with earlier
simulation work where different cosmological parameters were
assumed. We show this explicitly in Fig.~\ref{FigMczComp}, where we
compare the predictions of our model with the fitting formulae
proposed by \citet{Duffy2008}, \citet{Maccio2008} and \citet{Prada2012} 
for the WMAP5, WMAP3 and WMAP7 cosmologies, respectively. The dashed, dot-dashed 
and dashed lines show their fits, with thicker line type indicating the halo mass
range actually resolved by each simulation.
Our model is shown by the solid lines in Fig.~\ref{FigMczComp} and is
in good agreement with this earlier work. 

Fig.~\ref{FigMczComp} also makes clear that, at low masses, our
predicted $c(M)$ relation deviates systematically from a power law, in
a way that results in lower concentrations than expected from simply
extrapolating the power-law fits obtained at higher masses. With
hindsight, this is not surprising. The variance of the CDM power
spectrum varies slowly at very low masses, implying a weak mass
dependence of their formation histories and, therefore, similar
concentration. At the other end of the mass scale, a similar reasoning
explains why concentrations approach a constant value at very large
halo masses (see Sec.~\ref{SecMAHMprof}).

The weak mass dependence of concentration at low masses might lead to
important changes in the self-annihilation flux expected from low-mass
halos, and to modifications in the importance of the ``boost factor''
from substructure, which is expected to dominate the annihilation
luminosity \citep[e.g.][]{Kuhlen2008,Springel2008a}. We can estimate
the magnitude of the effect by integrating the annihilation luminosity
of a smooth NFW halo over the subhalo mass function $n(m_{\rm sub})$
from the free-streaming limit of the CDM particle ($\sim 10^{-6}\,
h^{-1} \, M_\odot$) up to the typical mass of the largest subhalo
($\sim 0.1\, M_0$). For $n(m_{\rm sub})\propto m_{\rm sub}^{-1.9}$
\citep{Springel2008b} and $M_0\sim 10^{12} {\rm M}_\odot$, we find
that a simple extrapolation of the power-law $c(M)$ proposed by Neto
et al.  results in a boost factor that exceeds by a factor of $\sim
20$ that predicted by our model \citep[see also][]{KVA2012}.  The
actual magnitude of the correction to the boost factor is not
straightforward to compute, however, since subhalo concentrations are
known to be affected by the tidal field of the main halo and the boost
depends in a complex manner on the subhalo-within-subhalo
hierarchy. We defer a detailed study of this issue to a future paper.

\section{Summary and Conclusions}
\label{SecConc}

We have used the Millennium Simulations, together with simulations of
the Aquarius Project, to investigate the dependence of halo
concentration on mass and redshift. This is a topic that has been
studied repeatedly through numerical simulations, but whose
understanding has so far eluded simple analytical modeling.

Our results confirm earlier claims that halo mass profiles are
strongly linked to the mass accretion history (MAH) of their main
progenitor and that the concentration of relaxed halos is solely a
function of the dimensionless mass parameter, $\nu(M,z)$
(eq.~(\ref{EqNu})). MAHs and mass profiles have similar shapes, and, for
given mass, each one can be described on average by a single
``concentration'' parameter. This implies that accretion histories may
be used to compute halo concentrations and vice versa. In practice, NFW
profiles may be fit to MAHs and their parameters used to predict the
mass profile concentration using eq.~(\ref{Eqcc}). Our analysis shows
that this procedure predicts accurately the mass-concentration
relation at all redshifts when MAHs are computed directly from
simulations.

We have also explored whether analytical MAH models can be used to
predict accurate concentrations. Although there are a number of MAH
models in the literature, some of them are not applicable to our
procedure, since they either adopt forms that do not depend solely on
$\nu$, or else predict MAHs that do not resemble NFW profiles. We find
that a model based on the framework proposed by
\citet{vandenBosch2002} predicts halo concentrations that agree with
simulations over a remarkable range of {\it ten decades} in halo
mass. It also predicts a dependence on cosmological parameters that
agrees with published simulations of different variants of the LCDM
cosmogony.

When extrapolated to low halos masses the model predicts
systematically lower concentrations than expected from the power-law
fits proposed in earlier work. This agrees with recent simulations of
microhalo formation, which find concentrations of order $\sim 60$-$80$
for $\sim 10^{-7} \, h^{-1} \, M_\odot$ halos at $z=0$
\citep{Anderhalden2013}. Although further work on the structure of
microhalos is clearly needed, our results make clear that care should
be taken when extrapolating the $c(M)$ relation to extremely low
masses, such as when computing the expected flux from
self-annihilation, or the ``boost factor'' from substructure.

The model we propose here resolves the long-standing difficulties that
have plagued earlier attempts to account analytically for the
evolution of the mass-concentration relation. It clarifies and extends
earlier models that link the concentration with particular features of
the mass accretion history, and allows the dependence of
concentrations on cosmological parameters to be estimated in a simple
way. We present a step-by-step description of how to
estimate the mass-concentration-redshift relation of LCDM halos in an
Appendix. This is a simple tool that should be of use when
interpreting observations that place constraints on the characteristic
density of the dark halos that surround galaxies and galaxy systems.

\section*{acknowledgements}
We wish to thank Gerard Lemson for useful discussions, as well as the
Virgo Consortium for access to the MS data. ADL acknowledges financial
support from the SFB (956) from the Deutsche
Forschungsgemeinschaft. V.S. acknowledges support by the DFG through
Transregio 33, ``The Dark Universe''. 
This work used the COSMA Data Centric system at Durham University, operated
 by the Institute for Computational Cosmology on behalf of the STFC DiRAC HPC
 Facility (www.dirac.ac.uk. This equipment was funded by a BIS National
 E-infrastructure capital grant ST/K00042X/1, DiRAC Operations grant
 ST/K003267/1 and Durham University. DiRAC is part of the National
 E-Infrastructure. We acknowledge partial financial support from 
by the National Science Foundation under Grant No. PHYS-1066293 and
the hospitality of the Aspen Center for Physics.

\bsp
\label{lastpage}

\appendix

\section{A model for computing the concentration-mass-redshift relation}
\label{AppvdB}

Our model is based on the model for accretion histories of
\citet{vandenBosch2002}. We provide here a simple summary of the basic
features needed to compute $c(M,z)$ for LCDM halos, and refer the
reader to the original paper for details.  In this model, the {\it
  average} MAH of a halo of mass $M_0$ at $z=0$ is approximated as
follows:
\begin{equation}
\log \frac{M(z)}{M_0/2}=\biggl(\frac{\log(1+z)}{\log(1+z_{f})}\biggr)^\chi,
\label{app:mah}
\end{equation}
where $z_f(M_0)$ is a characteristic ``half-mass'' formation time, and
$\chi$ is a parameter that depends on cosmology and mass,
\begin{eqnarray}
\label{nuB}
\lefteqn{\chi = 1.211 + 1.858 \, \log (1+z_f) + 
0.308 \, \Omega_{\Lambda}^{2} - } \nonumber \\
 & & 0.032 \, \log (M_0/[10^{11} h^{-1}  M_\odot]).
\end{eqnarray}
The ``formation redshift'', $z_f$, is given by 
\begin{equation}
  \delta_{\rm crit}(z_f)=\delta_{\rm crit}(0)+0.477\, \sqrt{2[\sigma^2(f\, M_0)-\sigma^2(M_0)]},
  \label{app:zfroot}
\end{equation}
where $\delta_{\rm crit}(z)=\delta_{\rm crit}^0/D(z)$
\citep{Lacey1993} and $f=0.068$ is a fitting parameter obtained
empirically from fits to the MS accretion histories. $D(z)$ is the
linear growth factor; and $\delta_{\rm crit}^0$ is the critical
density threshold for spherical collapse at $z=0$. The latter depends
(very weakly) on cosmology and can be accurately approximated by
$\delta_{\rm crit}^0=0.15\, (12\, \pi)^{2/3} \, \Omega_{\rm
  M}^\kappa$, with
\begin{equation}
\kappa=
\cases{
0     & if $\Omega_{\rm M}=1$ and $\Omega_\Lambda=0$,\cr
0.0185& if $\Omega_{\rm M}<1$ and $\Omega_\Lambda=0$,\cr
0.0055& if $\Omega_{\rm M}+\Omega_\Lambda=1$.\cr}
\label{app:alpha}
\end{equation}

This procedure fully specifies the assembly histories of halos
identified at $z=0$. We express them in terms of the dimensionless
mass parameter, $\nu=\delta_{\rm crit}/\sigma(M)$, and assume that this
is the only relevant dependency in order to compute mass accretion
histories for halos identified at any other redshift. 

These mass accretion histories are then expressed as $M(\rho_{\rm
  crit}(z))$ and fitted to the NFW mass-density profile given by
eq.~(\ref{EqNFW_enc}). The MAH concentration thus obtained is then
converted into a predicted mass profile concentration using
eq.~(\ref{Eqcc}).

\bibliographystyle{mn2e}
\bibliography{paper}

\end{document}